\documentclass[aps,pra,showpacs,twocolumn]{revtex4-1}
\usepackage{amsfonts}
\usepackage{amssymb}
\usepackage{amsmath}
\usepackage{graphicx}
\usepackage{epsfig}
\usepackage{color}
\usepackage{amsmath,bm}
\usepackage{booktabs}

\begin{document}

\title{Exact eigenstates with off-diagonal long-range order for interacting
bosonic systems}
\author{C. H. Zhang}
\author{Z. Song}
\email{songtc@nankai.edu.cn}

\begin{abstract}
Fermions and hardcore bosons share the same restriction: no more than one
particle can occupy a single site in a lattice system. Specifically, in one
dimension, two systems can share the same matrix representation. In this
work, we investigate both the fermion and hardcore-boson models with
nearest-neighbor (NN) interaction in a ring lattice. We construct the exact
eigenstates of the hardcore-boson model with resonant NN interaction\ and
show that they possess off-diagonal long-range order (ODLRO) in the
thermodynamic limit. In comparison, the fermionic counterpart does not
support such a feature due to the different particle statistics, although
they share an identical energy spectrum. In addition, we examine the effect
of the periodic boundary condition on the dynamics of the condensate states
through numerical simulations.
\end{abstract}

\maketitle

\affiliation{School of Physics, Nankai University, Tianjin 300071, China}

\section{Introduction}

Bose-Einstein condensation (BEC) represents a remarkable demonstration of
quantum phenomena manifesting on a macroscopic scale. \cite%
{bose1924plancks,einstein1924quantentheorie}. This unique state of matter
exemplifies the profound influence of quantum mechanics on the behavior of
matter at a level visible to the naked eye. It is characterized by the
emergence of a unified quantum state among a group of free bosons,
showcasing a fascinating synchronization in their behavior. This phenomenon
epitomizes the profound impact of quantum mechanics on the collective
properties of particles. Intuitively, one might expect that on-site
repulsive interactions would prevent the formation of BEC at moderate
particle densities, because fermions and hardcore bosons share the same
restriction: no more than one particle can occupy a single site in a lattice
system. \textcolor{red}{The one-dimensional interacting bosonic system has received
much attention from the very early years due to the special confinement in
real space. The origins of this interest can be traced back to the
Lieb-Liniger model, which provides a description of a one-dimensional gas
consisting of interacting particles. Characteristic of the interaction
between two bosons is the use of Dirac delta functions \cite{lieb1963exact}.
In the large-strength limit of the delta interaction, the Lieb-Liniger model
reduces to the Tonks-Girardeau gas \cite{girardeau1960relationship}. It is
not a Bose-Einstein condensate, as it does not exhibit any of the necessary
characteristics, such as off-diagonal long-range order or a unitary two-body
correlation function, even in the thermodynamic limit. The influence of the
delta-interaction strength on the off-diagonal long-range order (ODLRO) of
the ground state has been investigated \cite{colcelli2018deviations}. It
shows that the interaction can destroy the ODLRO. When the Tonks-Girardeau
gas is placed in a periodic potential, the lower energy physics can be
characterized by a tight-binding hardcore Bose Hubbard model, which
corresponds to the Hamiltonian in our work with zero NN interaction. This
model might be expected to be fermionic because the bosons cannot exchange
places. This peculiar feature enables us to map a non-interacting hardcore
boson chain to a free fermion system by Jordan-Wigner transformation, which
allows for an exact solution. Nevertheless, only numerical results of
correlation functions for finite systems can be obtained \cite%
{rigol2005ground,fu2006tonks}, rather than exact conclusions. }Furthermore,
significant effort has been dedicated to investigating and understanding the
impact of particle-particle interactions on the emergence of BEC \cite%
{shi1998finite,andersen2004theory,yukalov2004principal}.

Recent breakthroughs in the realm of cold atom experiments have paved the
way for a multifaceted platform that enables the exploration and realization
of diverse phases within both interacting and non-interacting bosonic
systems. This cutting-edge approach has significantly broadened our capacity
to study and manipulate quantum states of matter \cite%
{bloch2012quantum,atala2014observation,aidelsburger2015measuring,stuhl2015visualizing}
. Current experimental setups now enable the precise control of both
geometry and interactions, allowing for the direct investigation of the
real-time evolution of quantum many-body systems using engineered model
Hamiltonians. \cite{jane2003simulation,bloch2012quantum,blatt2012quantum}.
It thus enhances the theoretical predictions for exotic quantum phases in
interacting systems, potentially enabling their realization and experimental
testing. Exact solutions for quantum many-body systems are rare, yet they
are crucial for providing valuable insights into the characterization of new
forms of quantum matter and dynamic behaviors.

In this paper, we study one-dimensional extended Bose-Hubbard systems with
strong on-site interaction and its spinless fermionic counterpart. Under
certain conditions, the two systems are equivalent due to the mapping
between hardcore bosons and spinless fermions, as neither system allows more
than one particle to occupy a single site within a lattice. We construct the
exact eigenstates for the hardcore-boson model on a ring lattice with
resonant nearest-neighbor (NN) interactions and demonstrate that they
exhibit ODLRO in the thermodynamic limit. In comparison, we also study the
fermionic counterpart of the eigenstate and find that it does not support
long-range order due to the different particle statistics, even though it
shares an identical wave function structure. Furthermore, we investigate the
impact of boundary conditions on the stability of the long-range order in
the condensate state through numerical simulations for finite-size system of
quench dynamics. The finding not only reveals the possible condensation of
interacting bosons, but also clarifies the relationship between the hardcore
boson system and its fermion counterpart. \textcolor{red}{Here, we would like to
address the main points of the present work. (i) This work mainly focuses on
a hardcore boson tight binding ring with resonant NN interaction. We present
a set of exact eigenstates of the Hamiltonian, which are shown to have
ODLRO. It indicates that although the on-site interaction can disrupt the
ODLRO, the NN interaction can counteract this effect. (ii) It also provides
a set of exact eigenstates of the corresponding fermion Hamiltonian with
resonant NN interaction. Nevertheless, we show that such a set of fermionic
eigenstates does not have ODLRO.}

This paper is organized as follows. In Sec. \ref{Model Hamiltonians}, we
introduce two types of model Hamiltonians and explore the connections
between them. In Sec. \ref{Two-particle eigenstates}, we investigate simple
cases to introduce the main points of this work.\ In Sec. \ref{Hardcore
boson eigenstates with ODLRO}, we present the eigenstate exhibiting ODLRO.
In Sec. \ref{Fermionic counterpart}, we investigate the characteristics of
the eigenstate for the fermionic counterpart. In Sec. \ref{Stability of the
correlation function}, we conduct numerical simulations to demonstrate the
impact of boundary conditions on the stability of long-range order within
the condensate state. Finally, we present a summary of our results in Sec. %
\ref{Summary}.

\section{Model Hamiltonians}

\label{Model Hamiltonians}

We start our study from a general form of the Hamiltonian on a
one-dimensional lattice $N$ with periodic boundary condition%
\begin{equation}
H=-\kappa \sum_{l=1}^{N}a_{l}^{\dagger }a_{l+1}+\mathrm{H.c.}%
+V\sum_{l=1}^{N}n_{l}n_{l+1},  \label{H}
\end{equation}%
where $a_{l}^{\dagger }$ is the hardcore boson at the position $l$, and $%
n_{l}=a_{l}^{\dagger }a_{l}$. Here\ $\kappa $\ represents the hopping
strength across adjacent sites and $V$\ is the NN coupling constant. The
hardcore-boson operators satisfy the commutation relations%
\begin{equation}
\left\{ a_{l},a_{l}^{\dagger }\right\} =1,\left\{ a_{l},a_{l}\right\} =0,
\end{equation}%
and%
\begin{equation}
\left[ a_{j},a_{l}^{\dagger }\right] =0,\left[ a_{j},a_{l}\right] =0,
\end{equation}%
for $j\neq l$. The total particle number operator, $n=\sum_{l}n_{l}$, is a
conserved quantity because it commutes with the Hamiltonian, i.e., $\left[
n,H\right] =0$. Then one can investigate the system in each invariant
subspace with fixed particle number $n$. Under the open boundary condition,
we define $a_{N+1}=0$, while $a_{N+1}=a_{1}$ for the periodic boundary
condition.

One can perform the Jordan-Wigner transformation \cite{jordan1993paulische}%
\begin{equation}
a_{j}=\exp \left( i\pi \sum_{l<j}c_{l}^{\dagger }c_{l}\right) c_{j},
\label{JW transf}
\end{equation}%
to replace the hardcore boson operators by the fermionic operators $c_{j}$
and $c_{j}^{\dagger }$, which satisfy the commutation relations%
\begin{equation}
\left\{ c_{l},c_{l^{\prime }}^{\dagger }\right\} =\delta _{ll^{\prime
}},\left\{ c_{l},c_{l^{\prime }}\right\} =0.
\end{equation}%
The Hamiltonian is transformed to a spinless fermionic model%
\begin{eqnarray}
H_{\text{f}} &=&-\kappa \sum_{l=1}^{N-1}c_{j}^{\dagger }c_{j+1}+\kappa
(-1)^{n}c_{1}^{\dagger }c_{N}+\mathrm{H.c.} \\
&&+V\sum_{l=1}^{N}c_{j}^{\dagger }c_{j}c_{j+1}^{\dagger }c_{j+1},
\end{eqnarray}%
where $n$ is total number of the fermions (or hardcore bosons). \textcolor{red}{The
Hamiltonian $H_{\text{f}}$ is not a toy model and has received
much attention from the early years. It belongs to the class of physical
models that were proposed in cite \cite{hirsch1983phase,makhankov1984non}
some forty years ago. As a specific example, $H_{\text{f}}$ is
commonly referred to as the polaron model\ or the small polaron model, the
solution of which was investigated in many works \cite%
{pu1986exact,zhou1988infinite,zhou1990some,umeno1999fermionic} . Since that
time there have been several studies of the model. The solution of the
inverse problem\ for this model was formulated in \cite{gohmann2000solution}%
. A study of the fusion hierarchy is given in \cite{grabinski2013truncation}%
. And in addition, there have been numerous studies on the exact solution of
the model with various types of boundary conditions e.g. \cite%
{guan1999exact,xu2015small}. Apart from these, there are also some known results and existing body of literatures on the related model\cite{shastry1986infinite,sklyanin1988boundary,guan1998lax,zhou1988further,zhou1996boundary,micheletti1997complete,cao2013off,cao2013off2,cao2013off3,karaiskos2013bethe,li2014exact,cao2014spin,cao2015exact,zhang2014exact,hao2014exact,wang2015off,wu2022exact,li2024few}. In the following, we will show that
there exist a set of exact eigenstates of the Hamiltonian $H_{\text{f}}$ 
 with resonant NN interaction.}

Obviously, when considering an open chain system or in odd $n$ subspace, the
two Hamiltonians $H$ and $H_{\text{f}}$ have an identical matrix
representation and spectrum because particles cannot occupy the same site
simultaneously in either model. Accordingly, both Hamiltonians share the
same set of eigen wavefunctions but based on different bases. However, two
kinds of particles have different particle statistics. This should result in
distinct features between the two states, for fermion or hardcore boson,\
even though they share the same wavefunction for a properly chosen basis
set. We demonstrate this point by comparing the two-particle eigenstates of
two models in the next section.

\section{Two-particle eigenstates}

\label{Two-particle eigenstates}In this section, we start with simple cases,
in which the obtained results shed light on one of the main conclusions of
this work. Considering the Hamiltonian $H$ and $H_{\text{f}}$\ with zero $V$
and open boundary condition, a single-particle eigenstate with 
\begin{equation}
\varepsilon _{k_{n}}=-2\kappa \cos k_{n},
\end{equation}%
is given by 
\begin{equation}
\left\vert \psi _{k_{n}}\right\rangle =a_{k_{n}}^{\dagger }\left\vert
0\right\rangle ,
\end{equation}%
and%
\begin{equation}
\left\vert F_{k_{n}}\right\rangle =c_{k_{n}}^{\dagger }\left\vert
0\right\rangle ,
\end{equation}%
respectively. Here the quasi wave vector is defined as $k_{n}=n\pi /(N+1)$ ($%
n\in \lbrack 1,N]$), and the corresponding particle operators with mode $%
k_{n}$\ read%
\begin{equation}
a_{k_{n}}^{\dagger }=\sqrt{\frac{2}{N+1}}\sum_{l}\sin \left( k_{n}l\right)
a_{l}^{\dagger },
\end{equation}%
and%
\begin{equation}
c_{k_{n}}^{\dagger }=\sqrt{\frac{2}{N+1}}\sum_{l}\sin \left( k_{n}l\right)
c_{l}^{\dagger },
\end{equation}%
respectively. In this sense, all the eigenstates of $H_{\text{f}}$ can be
constructed using the set of operators $\left\{ c_{k_{n}}^{\dagger }\right\} 
$. Furthermore, all the eigenstates of $H$ can be derived from those of $H_{%
\text{f}}$ with the help of the Jordan-Wigner transformation.

For instance, the two-fermion ground state of $H_{\text{f}}$ has the form 
\begin{equation}
\left\vert F_{k_{1}k_{2}}\right\rangle =\frac{2}{N+1}\sum_{l,l^{\prime
}}\sin \left( k_{1}l\right) \sin \left( k_{2}l^{\prime }\right)
c_{l}^{\dagger }c_{l^{\prime }}^{\dagger }\left\vert 0\right\rangle ,
\end{equation}%
which is obviously a trivial state, describing two uncorrelated fermions.
Accordingly, the two-hardcore boson ground state of $H$ can be expressed as%
\begin{eqnarray}
\left\vert \psi _{k_{1}k_{2}}\right\rangle  &=&\frac{2}{N+1}%
\sum_{l<l^{\prime }}[\sin \left( k_{1}l\right) \sin \left( k_{2}l^{\prime
}\right)   \notag \\
&&-\sin \left( k_{1}l^{\prime }\right) \sin \left( k_{2}l\right)
]a_{l}^{\dagger }a_{l^{\prime }}^{\dagger }\left\vert 0\right\rangle ,
\end{eqnarray}%
which is obtained from $\left\vert F_{k_{1}k_{2}}\right\rangle $\ by using
the inverse Jordan-Wigner transformation. A question arises: whether the
state $\left\vert \psi _{k_{1}k_{2}}\right\rangle $ is still nontrivial. To
answer this question, we introduce a related free-boson model%
\begin{equation}
H_{\text{b}}=-\kappa \sum_{l=1}^{N-1}\left( b_{j}^{\dagger }b_{j+1}+\text{%
H.c.}\right) .
\end{equation}%
which is obtained by directly replacing $a_{l}$\ by the boson operator $b_{l}
$\ in $H$. The boson operators satisfy the commutation relations%
\begin{equation}
\left[ b_{j},b_{l}^{\dagger }\right] =\delta _{jl},\left[ b_{j},b_{l}\right]
=0,
\end{equation}%
and then $H_{\text{b}}$\ can be written as a diagonal form%
\begin{equation}
H_{\text{b}}=\sum_{n}\varepsilon _{k_{n}}b_{k_{n}}^{\dagger }b_{k_{n}},
\end{equation}%
where the canonical operator is defined as%
\begin{equation}
b_{k_{n}}^{\dagger }=\sqrt{\frac{2}{N+1}}\sum_{l}\sin \left( k_{n}l\right)
b_{l}^{\dagger }.
\end{equation}%
The pair-boson eigenstate or two-boson condensate state of mode $k_{n}$\ in
the large $N$ limit has the form%
\begin{eqnarray}
&&\left\vert \varphi _{k_{n}k_{n}}\right\rangle =\frac{1}{\sqrt{2}}%
b_{k_{n}}^{\dagger }b_{k_{n}}^{\dagger }\left\vert 0\right\rangle   \notag \\
&=&\frac{\sqrt{2}}{N+1}\sum_{l<l^{\prime }}2\sin \left( k_{n}l\right) \sin
\left( k_{n}l^{\prime }\right) b_{l}^{\dagger }b_{l^{\prime }}^{\dagger
}\left\vert 0\right\rangle .
\end{eqnarray}%
Here, we neglect the high-order infinitesimal terms on the\ components $%
\left( b_{l}^{\dagger }\right) ^{2}\left\vert 0\right\rangle $\ in the large 
$N$ limit. In this sense, state $\left\vert \varphi
_{k_{n}k_{n}}\right\rangle $\ can be regarded as a hardcore boson state.\
Then one can characterize the property of the hardcore state $\left\vert
\psi _{k_{1}k_{2}}\right\rangle $\ by the set of non-trivial basis $\left\{
\left\vert \varphi _{k_{n}k_{n}}\right\rangle \right\} $. Specifically, we
treat this task by calculating the quantity 
\begin{equation}
\gamma _{n}=\left\vert \langle \psi _{k_{1}k_{2}}\left\vert \varphi
_{k_{n}k_{n}}\right\rangle \right\vert .  \label{gamma}
\end{equation}%
The values of $\gamma _{l}$\ and $\sum_{n=1}^{l}\gamma _{n}^{2}$\ are listed
in Table I. \textcolor{red}{Here, the coefficient $\gamma _{l}$ represents the
amplitude of the overlap for two states $\left\vert \psi
_{k_{1},k_{2}}\right\rangle $ and $\left\vert \varphi
_{k_{n},k_{n}}\right\rangle $. We note that the state $\left\vert \varphi
_{k_{n},k_{n}}\right\rangle $ is the two-boson condensate state of mode $%
k_{n}$\ for the free boson model, while $\left\vert \psi
_{k_{1},k_{2}}\right\rangle $ is the two-particle ground state of the
hardcore boson chain. When the quantity $\sum \gamma _{l}$ approaches $1$,
the state $\left\vert \psi _{k_{1},k_{2}}\right\rangle $ becomes a
superposition of two-boson condensate state with different modes. In
contrast, the state $\left\vert F_{k_{1},k_{2}}\right\rangle $, as the
ground state of two-particle free fermion model, should no longer exhibit
any condensates. Therefore, the two states $\left\vert \psi
_{k_{1},k_{2}}\right\rangle $\ and $\left\vert F_{k_{1},k_{2}}\right\rangle $
possess distinct characteristics, although they are connected by the
Jordan-Wigner transformation. It is expected that such a conclusion can be
extended to the case with many particles. }However, further investigation of
the many-particle case along this line becomes complicated. We will turn to
the case with nonzero $V$, but for an excited eigenstate. In the following,
we will demonstrate that there exists an $n$-hardcore boson eigenstate for $H
$ with resonant $V$, which are condensed states with ODLRO.

\begin{table}[tbph]
	\caption{The values of $\gamma_l$ and $\sum_{n=1}^{l}\gamma_n^2$ in Eq. (\ref{gamma}) in the large N limit. It indicates that $\gamma_l$ decays as $l$ increases and states $\left\vert\varphi_{k_nk_n}\right\rangle$ are dominant in the state $\left\vert\psi_{k_1k_2}\right\rangle$. }
	\centering
	\renewcommand{\arraystretch}{1} 
	\begin{tabular}{ccccccccccccccccccccccccccccccc}
		\hline\hline
		$l$& & &1& & & &2& & & &3& & & &4\\
		\hline
		\rule{0pt}{20pt} 
		$\gamma_l$& & &\fontsize{12pt}{20pt} $\frac{256\sqrt{2}}{45\pi^2}$ & & & &\fontsize{12pt}{20pt}$\frac{1024\sqrt{2}}{315\pi^2}$& & & &\fontsize{12pt}{20pt}$\frac{256\sqrt{2}}{315\pi^2}$ & & & &\fontsize{12pt}{20pt}$\frac{4096\sqrt{2}}{10395\pi^2}$\\[5pt]
		$\sum_{n=1}^{l}\gamma_n^2$& & &0.664& & & &0.881& & & &0.895& & & &0.898\\[5pt]
		\hline\hline
		$l$& & &5& & & &6& & & &7& & & &8\\
		\hline
		\rule{0pt}{20pt}
		$\gamma_l$& & &\fontsize{12pt}{20pt} $\frac{6400\sqrt{2}}{27027\pi^2}$ & & & &\fontsize{12pt}{20pt}$\frac{1024\sqrt{2}}{6435\pi^2}$& & & &\fontsize{12pt}{20pt}$\frac{12544\sqrt{2}}{109395\pi^2}$& & & &\fontsize{12pt}{20pt}$\frac{16384\sqrt{2}}{188955\pi^2}$\\[5pt]
		$\sum_{n=1}^{l}\gamma_n^2$& & &0.899& & & &0.900& & & &0.900& & & &0.900 \\[5pt]
		\hline\hline   
	\end{tabular}%
\end{table}
\begin{figure}[htbp]
\centering
\includegraphics[width=0.5\textwidth]{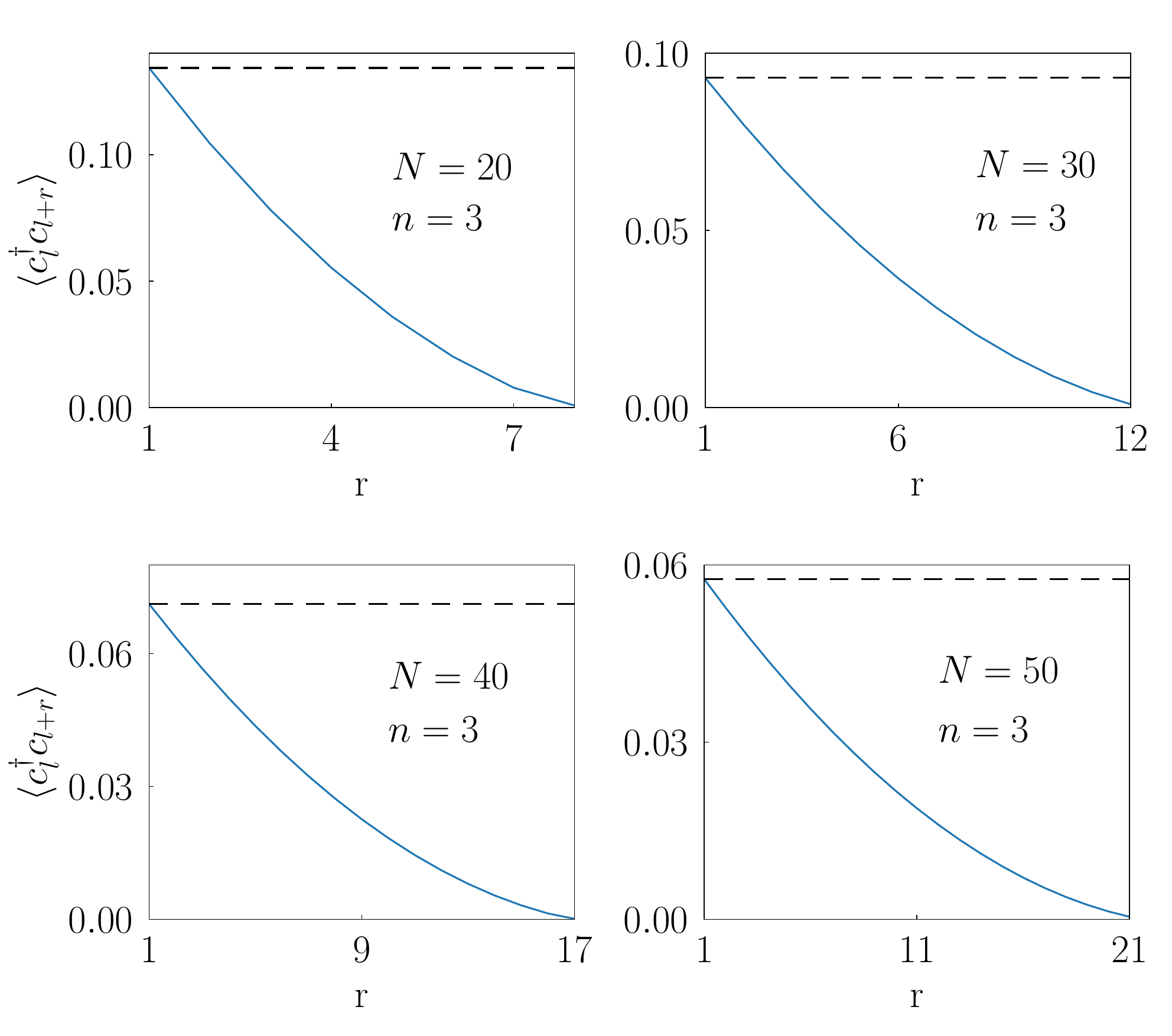}
\caption{Plots of the correlation function $\left\langle c_{l}^{\dagger
}c_{l+r}\right\rangle _{\text{f}}$ in Eq. (\protect\ref{CC_f}) for the
fermion counterpart of the hardcore boson condensate eigenstate obtained by
exact diagonalization for finite-size system. The system parameters $N$\ and 
$n$ are indicated in the panels and $q=2\protect\pi /N$. The dashed lines
indicate the value $\left( N-n\right) n/\left[ N(N-1)\right] $ from Eq. ( 
\protect\ref{<aa>}). As expected, the correlation function $\left\langle
c_{l}^{\dagger }c_{l+r}\right\rangle _{\text{f}}$ decays rapidly with
increasing $r$, exhibiting non--long-range correlations.}
\label{fig1}
\end{figure}
\begin{figure*}[htp]
\centering
\includegraphics[width=\textwidth]{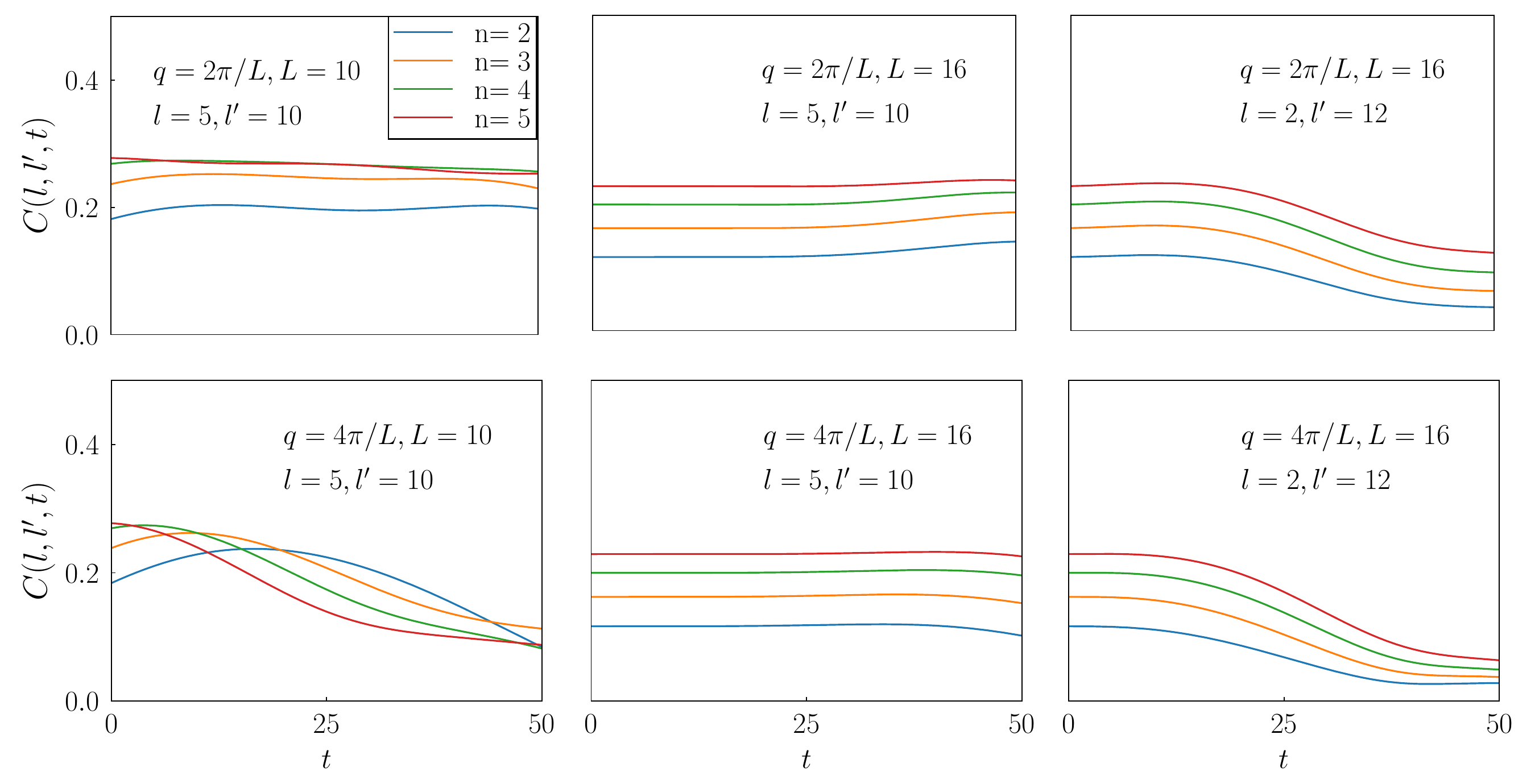}
\caption{Plots of the correlation function $C(l,l^{\prime },t)$ in Eq. ( 
\protect\ref{C(t)}) for the quench processes obtained by exact
diagonalization for finite-size system. The system parameters $N$, $q$, $n$, 
$l$ and $l^{\prime }$ are indicated in the panels. We can see that the
effect of the boundary condition becomes weaker for larger size $N$. The
correlation function varies slightly over time when the two related points, $%
l$ and $l^{\prime }$ are far from the boundaries.}
\label{fig2}
\end{figure*}

\section{Hardcore boson eigenstates with ODLRO}

\label{Hardcore boson eigenstates with ODLRO}

In this section, we focus on the hardcore boson\ Hamiltonian on a ring with
resonant NN interaction strength%
\begin{equation}
V=V_{\text{\textrm{r}}}=2\kappa \cos q,
\end{equation}%
where real number $q=2\pi m/N$ $(m=1,2,...,N)$ is the wave vector for the
ring.

In the following, we will show that state

\begin{equation}
\left\vert \psi _{n}\right\rangle =\frac{1}{\Omega _{n}}\left(
\sum_{l}a_{l}^{\dagger }e^{-iql}\right) ^{n}\left\vert 0\right\rangle ,
\end{equation}%
is an eigenstate of the system, where the normalization factor $1/\Omega
_{n} $ $=$ $n!\sqrt{C_{N}^{n}}$, and the vacuum state $\left\vert
0\right\rangle =\prod_{l}\left\vert 0\right\rangle _{l}$, with\ $%
a_{l}\left\vert 0\right\rangle _{l}=0$. To this end, we rewrite the
Hamiltonian in a specific form 
\begin{equation}
H=\sum_{l}\left( \mathcal{H}_{l}+V_{\text{\textrm{r}}}n_{l}\right) ,
\end{equation}%
that allows for easier analysis, where the sub-Hamiltonian i.e.,%
\begin{eqnarray}
&&\mathcal{H}_{l}=\kappa \lbrack a_{l}^{\dagger }a_{l+1}+\mathrm{H.c.}+2\cos
q(n_{l}n_{l+1}-n_{l}-n_{l+1})  \notag \\
&&+\left( e^{iq}n_{l}+e^{-iq}n_{l+1}\right) ],
\end{eqnarray}%
is non-Hermitian. Here, the introduced complex potentials are merely a
technical skill and do not involve non-Hermitian physics. State $\left\vert
\psi _{n}\right\rangle $\ can be shown to be the eigenstate of $H$, i.e., 
\begin{equation*}
H\left\vert \psi _{n}\right\rangle =nV_{\text{\textrm{r}}}\left\vert \psi
_{n}\right\rangle ,
\end{equation*}%
based on the following identities%
\begin{equation}
\left\{ 
\begin{array}{l}
\mathcal{H}_{l}[e^{-iql}a_{l}^{\dagger }+e^{-iq(l+1)}a_{l+1}^{\dagger
}]\left\vert 0\right\rangle _{l}\left\vert 0\right\rangle _{l+1}=0, \\ 
\mathcal{H}_{l}a_{l}^{\dagger }a_{l+1}^{\dagger }\left\vert 0\right\rangle
_{l}\left\vert 0\right\rangle _{l+1}=0, \\ 
\mathcal{H}_{l}\left\vert 0\right\rangle _{l}\left\vert 0\right\rangle
_{l+1}=0,%
\end{array}%
\right.  \label{3 identities}
\end{equation}%
which can be checked by a straightforward derivation.

In fact, state $\left\vert \psi _{n}\right\rangle $ can be expressed as the
form

\begin{equation}
\left\vert \psi _{n}\right\rangle =\frac{1}{\Omega _{n}}\left( [e^{-iql}
a_{l}^{\dagger }+e^{-iq(l+r)}a_{l+r}^{\dagger }]+D\right) ^{n}\left\vert
0\right\rangle ,
\end{equation}%
where operator%
\begin{equation}
D=\sum_{l^{\prime}}a_{l^{\prime}}^{\dagger
}e^{-iql^{\prime}}-[e^{-iql}a_{l}^{\dagger }+e^{-iq(l+r)}a_{l+r}^{\dagger }]
\end{equation}%
\ does not contain $a_{l}^{\dagger }$\ and $a_{l+r}^{\dagger }$. Here $r$ is
the arbitrary distance between two sites. We note that%
\begin{equation}
\lbrack e^{-iql}a_{l}^{\dagger }+e^{-iq(l+r)}a_{l+r}^{\dagger
}]^{2}\left\vert 0\right\rangle =2e^{-iq(2l+r)}a_{l}^{\dagger
}a_{l+r}^{\dagger }\left\vert 0\right\rangle ,
\end{equation}%
for hardcore boson operators and then we have%
\begin{equation}
\lbrack e^{-i\mathbf{q\cdot }l}a_{l}^{\dagger }+e^{-i\mathbf{q\cdot }%
(l+r)}a_{l+r}^{\dagger }]^{k}\left\vert 0\right\rangle =0,(k>2).
\end{equation}%
We reach the conclusion 
\begin{equation}
\mathcal{H}_{l}\left\vert \psi _{n}\right\rangle =0,
\label{Zero energy S eq}
\end{equation}%
by applying the identities in Eq. (\ref{3 identities}) on the state 
\begin{eqnarray}
\left\vert \psi _{n}\right\rangle &=&\frac{1}{\Omega _{n}}%
\{D^{n}+nD^{n-1}[e^{-iql}a_{l}^{\dagger }+e^{-iq(l+r)}a_{l+r}^{\dagger }] 
\notag \\
&&+e^{-iq(2l+r)}n(n-1)D^{n-2}a_{l}^{\dagger }a_{l+r}^{\dagger }\}\left\vert
0\right\rangle .
\end{eqnarray}%
with $r=1$.

Based on the above analysis, the correlation function can also be obtained as%
\begin{equation}
\left\langle \psi _{n}\right\vert a_{l}^{\dagger }a_{l+r}\left\vert \psi
_{n}\right\rangle =e^{-iqr}\frac{\left( N-n\right) n}{N(N-1)},  \label{<aa>}
\end{equation}%
from the relations

\begin{eqnarray}
&&a_{l}\left\vert \psi _{n}\right\rangle =\frac{1}{\Omega _{n}}%
\{nD^{n-1}e^{-iql}  \notag \\
&&+e^{-iq(2l+r)}n(n-1)D^{n-2}a_{l+r}^{\dagger }\}\left\vert 0\right\rangle ,
\notag \\
&&a_{l+r}\left\vert \psi _{n}\right\rangle =\frac{1}{\Omega _{n}}%
\{nD^{n-1}e^{-iq(l+r)}  \notag \\
&&+e^{-iq(2l+r)}n(n-1)D^{n-2}a_{l}^{\dagger }\}\left\vert 0\right\rangle .
\end{eqnarray}%
It indicates that state $\left\vert \psi _{n}\right\rangle $\ possesses
ODLRO according to ref. \cite{yang1962concept} due to the fact that the
correlation function does not decay as $r$\ increases.

To understand the features of the state $\left\vert \psi _{n}\right\rangle $%
, we consider the similarity between this state and the condensate state of
bosons $\left\vert \varphi _{q^{n}}\right\rangle $, which has the form%
\begin{equation}
\left\vert \varphi _{q^{n}}\right\rangle =\frac{1}{\sqrt{n!N^{n}}}\left(
\sum_{j}e^{iqj}b_{j}^{\dag }\right) ^{n}\left\vert 0\right\rangle .
\end{equation}%
We note that the state $\left\vert \psi _{n}\right\rangle $\ can be simply
obtained by projecting out the on-site multiple-occupied components from the
state $\left\vert \varphi _{q^{n}}\right\rangle $, i.e.,%
\begin{equation}
\left\vert \psi _{n}\right\rangle \propto P\left\vert \varphi
_{q^{n}}\right\rangle ,
\end{equation}%
where the operator $P$ acts such that%
\begin{equation}
P\left( b_{j}^{\dag }\right) ^{n}\left\vert 0\right\rangle =0,
\end{equation}%
for $n>1$. The corresponding correlation function $\left\langle \varphi
_{q^{n}}\right\vert b_{l}^{\dag }b_{l+r}\left\vert \varphi
_{q^{n}}\right\rangle $ can be derived analytically. In fact, based on the
relations%
\begin{equation}
b_{l}\left\vert \varphi _{q^{n}}\right\rangle =\frac{1}{\sqrt{n!N}}%
e^{-iql}b_{q}\left( b_{q}^{\dag }\right) ^{n}\left\vert 0\right\rangle ,
\end{equation}%
and%
\begin{equation}
b_{l+r}\left\vert \varphi _{q^{n}}\right\rangle =\frac{1}{\sqrt{n!N}}%
e^{-iq\left( l+r\right) }b_{q}\left( b_{q}^{\dag }\right) ^{n}\left\vert
0\right\rangle ,
\end{equation}%
a straightforward derivation reveals that the correlation function%
\begin{equation}
\left\langle \varphi _{q^{n}}\right\vert b_{l}^{\dag }b_{l+r}\left\vert
\varphi _{q^{n}}\right\rangle =e^{-iqr}\frac{n}{N},  \label{<bb>}
\end{equation}%
is a finite number, thereby indicating the presence of ODLRO.

Comparing Eq. (\ref{<aa>}) and (\ref{<bb>}), we note that the two
correlation functions, $\left\langle \psi _{n}\right\vert a_{l}^{\dagger
}a_{l+r}\left\vert \psi _{n}\right\rangle $ and $\left\langle \varphi
_{q^{n}}\right\vert b_{l}^{\dag }b_{l+r}\left\vert \varphi
_{q^{n}}\right\rangle $, are identical in the dilute density limit $n/N\ll 1$%
\ and show no qualitative difference for finite density $n/N$. It indicates
that the two states $\left\vert \varphi _{q^{n}}\right\rangle $\ and $%
\left\vert \psi _{n}\right\rangle $\ share the same feature as a condensate
state. \textcolor{red}{In addition, this conclusion can be verified in another way 
\cite{colcelli2018deviations}. In fact, the largest eigenvalue $\lambda _{0}
$ of one-body-density matrix $\rho $ can be evaluated as%
\begin{equation}
\lambda _{0}=\frac{Nn-n^{2}+N}{N}\approx \frac{(N-n)n}{N}.
\end{equation}%
Here the matrix is defined as %
\begin{equation}
\rho _{ij}=\left\langle \psi _{n}\right\vert a_{i}^{\dagger }a_{j}\left\vert
\psi _{n}\right\rangle ,
\end{equation}%
for the state $\left\vert \psi _{n}\right\rangle $. 
We note that $\lambda _{0}\sim n$, which means the
existence of ODLRO}.

\section{Fermionic counterpart}

\label{Fermionic counterpart}

Now, let us turn to the discussion of the fermionic Hamiltonian $H_{\text{f}%
} $. When considering the hardcore boson Hamiltonian $H$ on a ring lattice
and within an invariant subspace with an odd particle number $n$, the
hardcore-boson state $\left\vert \psi _{n}\right\rangle $ should have its
fermionic counterpart $\left\vert F_{n}\right\rangle $: the eigenstate of
the fermionic Hamiltonian $H_{\text{f}}$ with $V=V_{\text{\textrm{r}}}$.
Taking the Jordan-Wigner transformation, we have%
\begin{equation}
\left\vert F_{n}\right\rangle =\frac{1}{\Omega _{n}}[e^{-iqj}\sum_{j}\exp
(i\pi \sum_{l<j}c_{l}^{\dagger }c_{l})c_{j}^{\dagger }]^{n}\left\vert
0\right\rangle .
\end{equation}%
Although this eigenstate has an analytical expression, its features are
still subtle. A natural question arises: whether such a state inherits the
condensate feature from the state $\left\vert \psi _{n}\right\rangle $.
Based on the above analysis, the mathematical identity%
\begin{eqnarray}
&&\left\langle F_{n}\right\vert \exp (i\pi \sum_{l<j}c_{l}^{\dagger
}c_{l})c_{j}^{\dagger }\exp (i\pi \sum_{l<j}c_{l}^{\dagger
}c_{l})c_{j+r}\left\vert F_{n}\right\rangle  \notag \\
&=&e^{-iqr}\frac{\left( N-n\right) n}{N(N-1)},
\end{eqnarray}%
can be derived, indicating a weird correlation function.\ Nevertheless, the
physical implications of the correlation remain unclear at this time. On the
other hand, deriving the expression of the conventional correlation function 
\begin{equation}
\left\langle c_{l}^{\dagger }c_{l+r}\right\rangle _{\text{f}}=\left\vert
\left\langle F_{n}\right\vert c_{l}^{\dagger }c_{l+r}\left\vert
F_{n}\right\rangle \right\vert  \label{CC_f}
\end{equation}%
is somewhat challenging, which is believed to characterize the property of
the state $\left\vert F_{n}\right\rangle $. Numerical simulations of the
correlation function $\left\langle c_{l}^{\dagger }c_{l+r}\right\rangle _{%
\text{f}}$ for a fermionic system with odd $n$ are performed. The eigenstate 
$\left\vert F_{n}\right\rangle $\ can be obtained by numerically exact
diagonalization in the odd-$n$\ invariant subspace of finite-sized systems.
The results for systems with representative parameters are presented in Fig. %
\ref{fig1}. We can see that the correlation function demonstrates power-law
decay, indicating non-long range order. Together with the results for
two-particle ground states in the previous section, we find that the
long-range order in a hardcore boson state may be destroyed by the
Jordan-Wigner transformation and does not appear in its Fermionic
counterpart.
\begin{figure*}[htp]
	\centering
	\includegraphics[width=\textwidth]{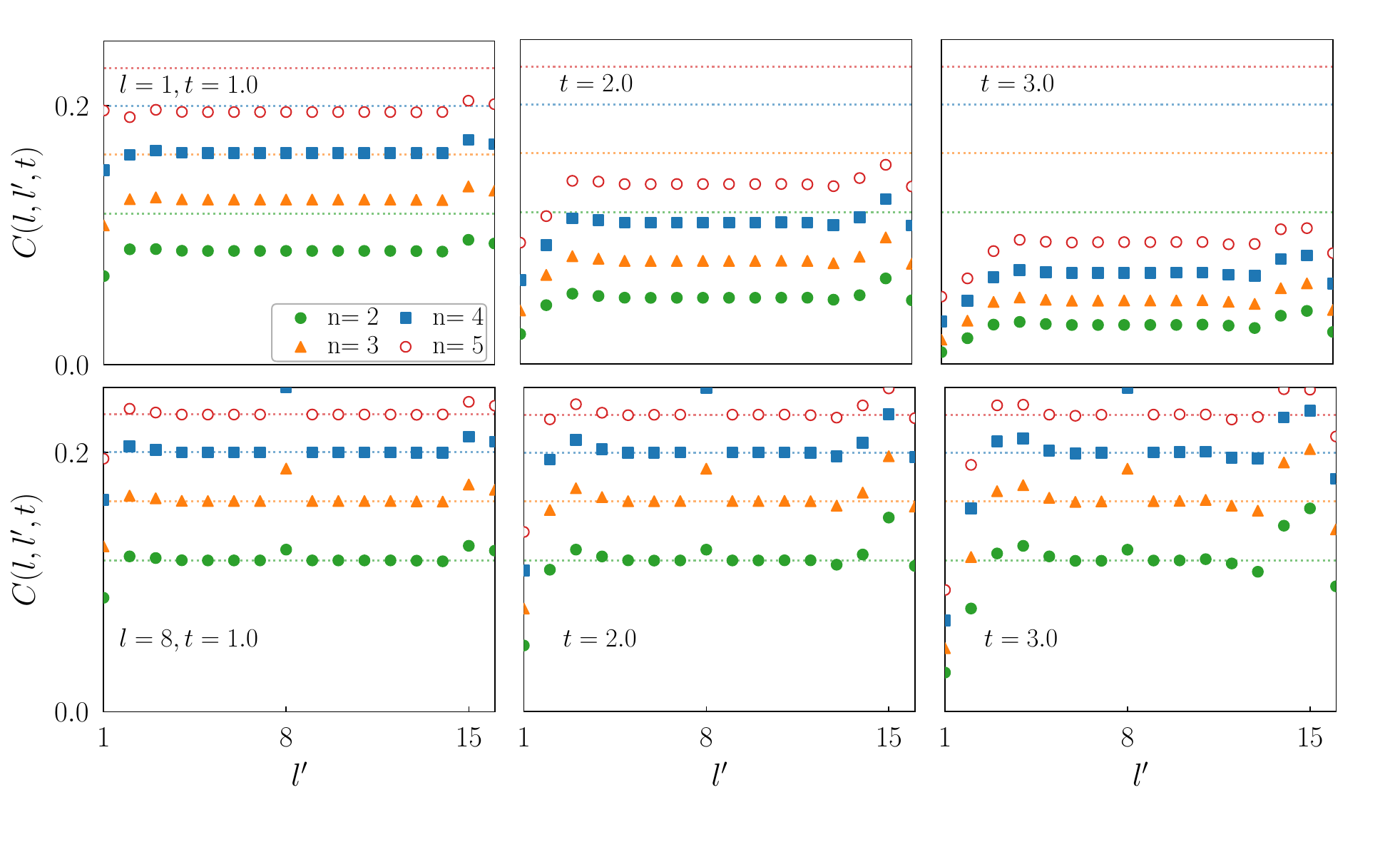}
	\caption{Snapshots of the correlation function $C(l,l^{\prime },t)$ in Eq.
		(46) for different $t$. The parameters $n$, $l$ and $l^{\prime }$ are indicated
		in the panels and $N=16$, $q=2\protect\pi/N$. The dashed lines represent the
		values in Eq. (32) where the parameter $n$ matches with numerical results in
		the same color. For $l$ in the bulk, the correlation function is
		stable through the evolution and corresponds with the dashed line, while
		for $l$ at the boundary, the correlation functions are consistent for $%
		l^{\prime}$ in the bulk but decay through the evolution and deviate from the dashed
		line. The results for $l=l^{\prime }$ can be ignored because it
		represents $\left\langle n_{j}\right\rangle $, the average particle number
		on site j.}
	\label{fig3}
\end{figure*}
\section{Stability of the correlation function}
 \label{Stability of the correlation function} It seems
that our conclusion depends on the periodic boundary conditions. The
condensate state $\left\vert \psi _{n}\right\rangle $\ is an exact
eigenstate of the Hamiltonian on a ring. However, the open boundary
condition may lead to a deviation of the eigenstate from the expression of $%
\left\vert \psi _{n}\right\rangle $\ and\ is usually taken in practice. In
this section, we focus on the influence of the boundary condition on the
existence of the long-range order condensate state. Our strategy is to
examine the dynamic response of the correlation function under a quenching
process. Specifically, we numerically compute the time evolution of the
state $\left\vert \phi \left( 0\right) \right\rangle =\left\vert \psi
_{n}\right\rangle $\ as an initial state under the same Hamiltonian with
open boundary conditions. The evolved state can be expressed as
\begin{equation} \left\vert \phi \left( t\right) \right\rangle =e^{-iH_{\text{%
			\textrm{OBC}}}t}\left\vert \psi _{n}\right\rangle , 
\end{equation}
where $H_{%
\text{\textrm{OBC}}}$\ is the Hamiltonian with open boundary condition. The
evolved state is calculated by exact diagonalization for finite systems\
with several typical particle filling number $n$. We focus on the
correlation function
\begin{equation} C(l,l^{\prime },t)=\left\langle \phi
\left( t\right) \right\vert a_{l}^{\dagger }a_{l^{\prime }}\left\vert \phi
\left( t\right) \right\rangle , \label{C(t)} 
\end{equation}
which measures the nature of condensation. \textcolor{red}{We plot $C(l,l^{\prime },t)${\ in Fig. \ref{fig2} as function of }${t}$
for selected systems and particle numbers. Here, we compute the correlation functions for three distinct cases of positions $l$ and $
l^{\prime }$: (i) one of $l$ or $l^{\prime }$\ is located at the end of the
chain; (ii) both $l$ and $l^{\prime }$\ are located in the bulk of the
chain; (iii) both $l$ and $l^{\prime }$\ are located at the ends of the
chain. The results indicate that the correlation functions in case (ii)
remain stable during the quenching dynamics. This suggests that boundary
conditions do not significantly influence the correlations in the bulk.
Consequently, the conclusions drawn for a ring system can be extended to a
chain system in the thermodynamic limit. Therefore, we conclude that there exists an excited
state with ODLRO in the hardcore boson model on an infinite-size chain
lattice.\\
For a clear observation of ODLRO, we also plot the snapshots of $C(l,l^{\prime },t)${\ in Fig. \ref{fig3} as function of }$l^{\prime}$ and compare the result with the analytical result in Eq. (32). The dashed lines indicate the
analytical values and the parameter $n$ matches with the numerical results in the same color. As shown in figure,
the results remain stable through the evolution and correspond with the analytical results
when $l$ and $l^{\prime}$ are located in the bulk, as the
conclusion mentioned above. And for the condition $l$ located at the
boundary, the correlation functions are still consistent for the condition $l^{\prime}$ in the bulk but
decay through the evolution and deviate from the dashed line, which means the
stability of correlations at the boundary is influenced. The results for
$l=l^{\prime }$ can be ignored because it represents $\left\langle
n_{j}\right\rangle $, the average particle number on site j.}

\section{Summary} \label{Summary} In summary, we have studied the extended hardcore
Bose-Hubbard model on one-dimensional\ lattices. A set of exact eigenstates
are constructed and have the following implications: (i) Strong on-site
repulsion and NN interaction cannot block the formation of a BEC under
moderate particle density, when the interacting strengths are deliberately
assigned. This conclusion can be extended to high dimensional systems.\ (ii)
Although there is an equivalent fermionic system derived from the
Jordan-Wigner transformation, its corresponding eigenstate does not exhibit
long-range order. (iii) The exactness of our results hinges on the periodic
boundary conditions. Moreover, numerical simulations demonstrate that the
long-range order remains robust even under open boundary conditions. Our
findings provide solid evidence for the existence of BECs for interacting
bosonic systems. \section*{Acknowldgement} This work was supported by the
National Natural Science Foundation of China (under Grant No. 12374461).
\bibliographystyle{unsrt} \bibliography{reference2} 
\end{document}